# Angular distribution of high power radiation from a charge rotating around a dielectric ball

L.Sh. Grigoryan[*], A.A. Saharian, H.F. Khachatryan, M.L. Grigoryan,
A.V. Sargsyan and T.A. Petrosyan

*Institute of Applied Problems of Physics NAS RA,
25 Hr. Nersessian Str., 0014 Yerevan, Republic of Armenia
E-mail*: levonshg@mail.ru

ABSTRACT: We study the angular distribution of the radiation from a relativistic charged particle uniformly rotating along an equatorial orbit around a dielectric ball. Earlier it was shown that for some values of the problem parameters and in the case of weak absorption in the ball material, the radiation intensity on a given harmonic can be essentially larger than that for the same charge rotating in the vacuum or in a homogeneous transparent medium having the same real part of dielectric permittivity as the ball material. The generation of such high power radiation is a consequence of the constructive superposition of electromagnetic oscillations of Cherenkov radiation induced near the trajectory of the particle and partially locked inside the ball. The angular distribution of the number of the emitted quanta is investigated for such high power radiation. It is shown that the radiation is mainly located in the angular range near the rotation plane determined by the Cherenkov condition for the velocity of the charge projection on the ball surface. The numerical analysis is given for balls made of strontium titanate, melted quartz and teflon in the gigahertz and terahertz frequency ranges.

---

[*]Corresponding author.

# Contents

## 1. Introduction

Due to unique properties such as the high intensity, high degree of collimation, and wide spectral range, synchrotron radiation (SR) (see, e.g., [1,2]) serves as an extraordinary research tool for advanced studies in both the fundamental and applied sciences. The wide applications of SR motivate the investigations of various mechanisms for the control of the spectral and angular characteristics. In particular, by taking into account that the characteristics of high-energy electromagnetic processes can be essentially changed in the presence of matter, it is of interest to study the influence of media on the properties of SR. Already in the case of a homogeneous medium the interplay of SR and Cherenkov radiation leads to remarkable effects [3-9].

    Additional possibilities for the control of SR parameters appear in inhomogeneous media. In particular, the interfaces separating two media with different electromagnetic properties can be used to control the radiation flow. As examples of this kind of exactly solvable problems, in a series of papers initiated in [10-12] we have considered SR in spherically and cylindrically symmetric layered media. It has been shown that the presence of boundaries leads to additional interesting features absent in the case of a homogeneous media. In particular, the investigations of SR from a charge rotating along an equatorial orbit around/inside a dielectric ball [13,14] showed that, when the Cherenkov condition for the ball material and particle speed is satisfied, the particle may generate and emit to outer space radiation field quanta (so called "resonant" radiation), the number of which exceeds the corresponding value for the case of a homogeneous and transparent medium in several dozens of times (see also [15] and references therein). Similar features take place for the angular distribution of SR in the case of cylindrical symmetry (see [16-18] and references therein).

    In this paper we investigate the angular distribution of the "resonant" radiation from a charged particle rotating around a dielectric ball in the gigahertz and terahertz frequency ranges. The numerical examples are given for balls made of strontium titanate, melted quartz or teflon.

## 2. Problem setup and the number of radiated quanta

We consider the uniform rotation of a relativistic electron with velocity $v$ along circular orbit of radius $r_e$ in the equatorial plane of a dielectric ball in the vacuum. The dielectric permittivity is written as a step function $\varepsilon(r) = \varepsilon_b + (1-\varepsilon_b)\Theta(r-r_b)$, where $r_b$ is the radius of ball, and $\varepsilon_b = \varepsilon_b' + i\varepsilon_b''$ is the complex valued permittivity for the ball material. Here a spherical coordinate system $r, \theta, \varphi$ is chosen with the origin at the centre of the ball. The polar axis is perpendicular to the equatorial plane of the electron rotation. The magnetic permeability of the medium is taken to be 1. We assume that the back reaction of the radiation on electron is small and can be neglected or it is compensated by an external field. Note that the circular motion of electrons with energies of the order of 1 MeV (below we will consider energies $\leq 2$ MeV) for the orbit radius of the order 1 cm can be generated by magnetic fields having the order 0.3 tesla. Magnetic fields with this order of strength are available in laboratories. For example, the magnetic fields in gyrotrons can be larger by an order of magnitude. The results given below are valid for an electron bunch as well replacing the electron charge by the bunch charge, under the assumption that the bunch length is smaller than the radiation wavelength.

The electron radiates on discrete angular frequencies $\omega_k = kv/r_e$ with the harmonic number $k = 1,2,3...$. We denote by $W_{kT}$ the energy radiated on a given harmonic $k$ per period $T = 2\pi r_e / v$ of electron rotation. For the related number of the radiated quanta one has

$$N_k = W_{kT} / \hbar \omega_k. \qquad (2.1)$$

The angular distribution $n_k(\theta)$ of the number of emitted quanta on a given harmonic $k$ is defined in accordance with the relation

$$N_k = \int_0^\pi n_k(\theta) d\theta, \qquad (2.2)$$

where $\theta$ is the corresponding polar angle.

In a homogeneous transparent medium with dielectric permittivity $\varepsilon$ (the medium fills whole space) the expressions for the total number of quanta radiated on a given harmonic and for its angular distribution read (see the corresponding equations (111) and (110) in [8] for the radiation intensity)

$$N_k^{(0)}(\varepsilon) = \frac{N_0}{\beta\sqrt{\varepsilon}}[2\beta^2 J'_{2k}(2k\beta) + (\beta^2 - 1)\int_0^{2k\beta} J_{2k}(x)dx], \qquad (2.3)$$

$$n_k^{(0)}(\varepsilon, \theta) = \frac{N_0 \cdot k}{\sqrt{\varepsilon}}\sin\theta \cdot [\cot^2(\theta)J_k^2(k\beta\sin\theta) + \beta^2 J'^2_k(k\beta\sin\theta)], \qquad (2.4)$$

where $N_0 = 2\pi e^2 / \hbar c \approx 0.0459$, $\beta = v\sqrt{\varepsilon}/c$ and $J_k(x)$ is the Bessel function. The special case $\varepsilon = 1$ of these formulae corresponds to SR in vacuum (see, e.g., [1]).

For an electron rotating around a dielectric ball, in [11,13] the expression

$$N_k = \frac{2N_0}{k}\sum_{s=0}^\infty (|a_{kE}(s)|^2 + |a_{kH}(s)|^2), \qquad (2.5)$$

was derived for the number of quanta emitted per rotation period. Here $a_{kE}(s)$ and $a_{kH}(s)$ describe the contributions of the electric and magnetic multipoles respectively. The corresponding expressions are given in [13] and they are valid without any restriction on the dispersion law for dielectric permittivity. The angular distribution of the radiated quanta is determined by the formula

$$n_k(\theta) = \frac{16\pi^2 N_0}{k}\sin\theta \cdot \left|\sum_{s=0}^\infty (-1)^s [a_{kE}(s)\vec{X}^{(2)}_{k+2s,k}(\theta,0) + ia_{kH}(s)\vec{X}^{(3)}_{k+2s+1,k}(\theta,0)]\right|^2, \qquad (2.6)$$

where $\vec{X}_{l,m}^{(\mu)}(\theta,\varphi)$ are spherical vectors of the electric ($\mu=2$) and magnetic ($\mu=3$) types [10,19]. Equation (2.6) is obtained from formula (19.1) in [11] after simple transformations and with notations introduced in [13]. In the absence of dielectric ball ($\varepsilon_b=1$) the calculations on the base of formulae (2.5), (2.6) give the same results as those obtained using the synchrotron radiation theory well-known formulae (2.3), (2.4) with $\varepsilon=1$.

## 3. Numerical analysis

Numerical calculations using formulae (2.5), (2.6) were carried out for three types of dielectric balls made of strontium titanate, melted quartz and teflon with small energy losses for electromagnetic waves in the gigahertz and terahertz frequency ranges. For strontium titanate $\varepsilon_b = \varepsilon_b' + i\varepsilon_b'' = 231(1+0.00019i)$ near the frequency 1GHz, for melted quartz $\varepsilon_b = 3.78(1+0.0001i)$ at the frequency of 10 GHz and for teflon $\varepsilon_b = 2.2(1+0.0002i)$ at the frequency of 600 GHz [20]. In figure 1 the results of numerical calculations for an electron with the energy of $E_e = 0.54\,\text{MeV}$ rotating in an orbit with the radius of $r_e \approx 1.54\,\text{cm}$ around a dielectric ball made of strontium titanate are presented. The radiation on the first harmonic $k=1$ is considered with frequency $\omega_k/2\pi = 1\,\text{GHz}$.

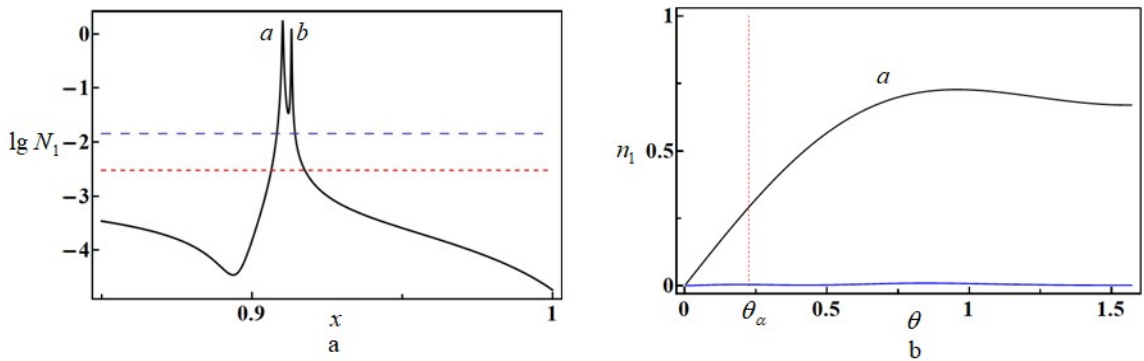

**Figure 1.** (a) The number $N_k$ of electromagnetic field quanta on the harmonic $k=1$, generated per rotation period of electron around a ball made of strontium titanate, versus the ratio $x = r_b/r_e$. (b) The corresponding angular distribution $n_k(\theta)$ for $x = x_a = 0.91024$ (for details see the text).

In figure 1(a) we have presented the number $N_1$ of electromagnetic field quanta generated per rotation period of electron around a dielectric ball versus the ratio $x = r_b/r_e$ of the radius of ball and that of electron orbit. In the absence of the ball (SR in the vacuum) one has $N_1^{(0)}(\varepsilon)|_{\varepsilon=1} \approx 0.003$ (see (2.3), the lower horizontal dashed line in figure 1(a)). According to figure 1(a), $N_1 \approx N_1^{(0)}(\varepsilon)|_{\varepsilon=1}$ practically for all values of $x$ except for $0.9 < x < 0.92$. In that region there are strongly expressed peaks $a,b$.

According to (2.3), for an electron with the values of the parameters $E_e$ and $r_e$ given above and circulating in an infinite, homogeneous and transparent medium with $\varepsilon = 231$, the number of quanta emitted on the first harmonic would be $N_1^{(0)}(\varepsilon) \approx 0.0143$ (the upper horizontal dashed line in figure 1(a)). For the higher peak in figure 1(a) one has $x_a = 0.91024$, $N_1 \approx 1.78$. Here the amplification of $N_1$ more than 100 times (more precisely $N_1/N_1^{(0)}(\varepsilon) \approx 125$) is due to the fact that electromagnetic oscillations of Cherenkov radiation induced along the trajectory of particle are partially locked inside the ball and superimposed in nondestructive way. Such a constructive superposition of electromagnetic oscillations (resonance) is accompanied by a powerful radiation at large distances from the system. Similar results (see, e.g., figures 2(a), 3(a)) are obtained for a number of other sets for the values of the parameters. As seen from figure 1(a), the radiation in a homogenous medium is stronger than that in the vacuum. Here and in what follows we compare with the radiation in a homogeneous medium in order to show that even in that case the effect induced by the ball is essentially stronger. For the radiation in a homogeneous medium we could consider also a weak absorption. In this case an additional constraint should be imposed on the observation point. The distance from the charge must be smaller than the characteristic length of the radiation absorption and the results will not be changed significantly.

In figure 1(b) (see also figures 2(b), 3(b)) we display the angular distribution of the radiation generated per period of the rotation of a charged particle around a dielectric ball. The calculations were carried out by formula (2.6). The lower curves (in figures 1(b),2(b),3(b)) correspond to the case of rotation of an electron in a homogeneous and transparent infinite medium with $\varepsilon = \varepsilon'_b$. The vertical dashed line marks the angle $\theta_a = \pi/2 - \theta_{Ch}$ corresponding to the emission angle of Cherenkov radiation from a superluminal electron moving rectilinearly with velocity $v_* = r_b v/r_e$ in a homogeneous and transparent dielectric medium having permittivity $\varepsilon = \varepsilon'_b$. Here the angle $\theta_{Ch}$ is determined from the Cherenkov condition

$$\cos\theta_{Ch} = c/v_*\sqrt{\varepsilon'_b}.  \qquad (3.1)$$

Note that $v_*$ is the velocity of the circulating electron projection on the ball surface.

From the data presented in figures 1(b), 2(b), 3(b) it follows that (i) the radiation is mainly located in the angular range

$$\theta \in [\theta_a, \pi - \theta_a] = [\pi/2 - \theta_{Ch}, \pi/2 + \theta_{Ch}], \qquad \varphi \in [0, 2\pi], \qquad (3.2)$$

and (ii) the specific form of the angular distribution of the number of emitted quanta is determined by the values of the parameters $k, v, \varepsilon_b, r_b/r_e$.

In figure 2, we present the results of numerical calculations for an electron with the energy $E_e = 2\,\text{MeV}$ rotating along the orbit with the radius $r_e \approx 3.69\,\text{cm}$ around a ball made of melted quartz. The radiation on the harmonic $k = 8$ is considered corresponding to frequency $\omega_k/2\pi = 10\,\text{GHz}$ (for data presented in figure 2(a) see also [13]).

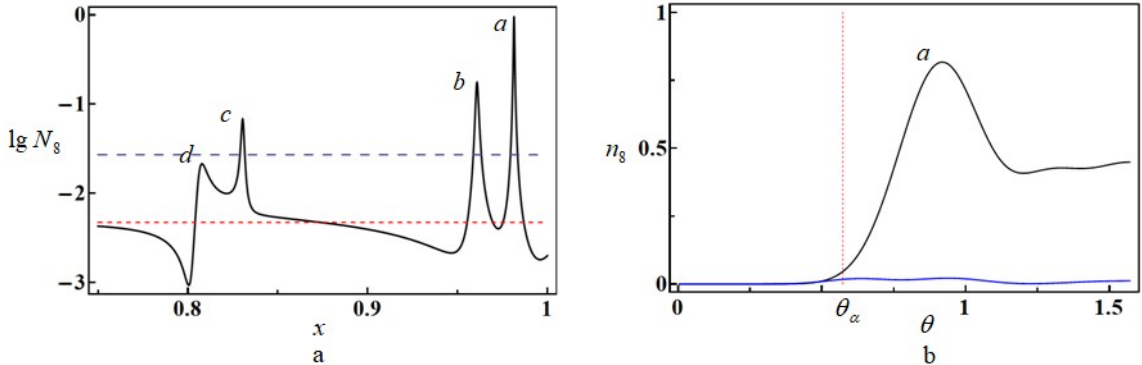

**Figure 2.** (a) The same as in figure 1(a) for a ball made of melted quartz and for $k = 8$ [13].
(b) The same as in figure 1(b) for a ball of melted quartz and for $k = 8$, $x = x_a = 0.9815$.

As in the case of strontium titanate $N_8 \approx N_8^{(0)}(\varepsilon)|_{\varepsilon=1} \approx 0.00475$ (see the lower dotted line in figure 2(a)) for almost all $x$ except $0.8 < x < 0.85$ and $0.95 < x < 1$. In those regions peaks appear. For the highest one we have $x_a = 0.9815$, $N_8 \approx 0.951$, and $N_8 / N_8^{(0)}(\varepsilon) \approx 35$, where $N_8^{(0)}(\varepsilon) \approx 0.0274$ (the upper dotted line in figure 2(a)). The corresponding value of the radius of the ball $r_b \approx 3.62$ cm (see also [13]). In figure 3 we present the results of similar numerical calculations for an electron with the energy of $E_e = 2$ MeV rotating in an orbit with the radius $r_e \approx 0.154$ cm around a dielectric ball made of teflon. The radiation is considered on the harmonic $k = 20$ with the frequency $\omega_k / 2\pi = 600$ GHz.

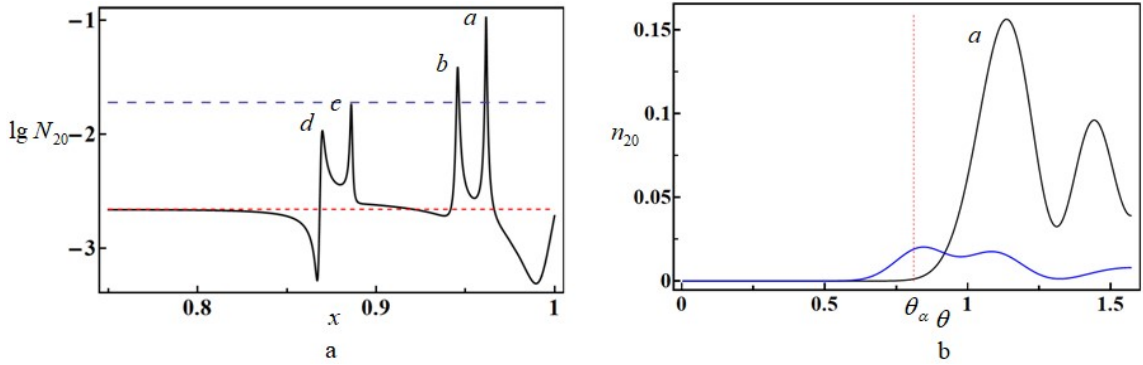

**Figure 3.** (a) The same as in figure 1(a) for the case of a ball made of teflon and for $k = 20$.
(b) The same as in figure 1(b) for a ball made of teflon and for $k = 20$, $x = x_a = 0.9616$.

As in the previous cases, in some regions peaks appear. For the highest of them one has the parameters $x = x_a = 0.9616$, $N_{20} \approx 0.107$ and $r_b \approx 0.148$ cm with $N_{20} / N_{20}^{(0)}(\varepsilon) \approx 5$, where $N_{20}^{(0)}(\varepsilon) \approx 0.019$ (the upper dotted line in figure 3(a)).

## 4. Conclusions

We have investigated the angular distribution of the radiation of a relativistic electron circulating around a dielectric ball. Earlier it was shown that for some values of the problem parameters on some harmonics, the electron can generate several times more electromagnetic field quanta than in the case of an electron rotating in a continuous, infinite and transparent medium having the same real part of the permittivity as the substance of the ball. The angular distribution of such powerful ("resonant") radiation, averaged over the rotation period, is determined by formula (2.6). That radiation is mainly located in the angular range (3.2) determined by the Cherenkov angle (3.1). This shows the relation of the high power radiation outside the ball with the Cherenkov radiation generated inside the ball. The specific form of the angular distribution in the range (3.2) depends on the radiation harmonic and on the values of the parameters $v, \varepsilon_b, r_b / r_e$.

Note that in a number of free electron based sources of THz radiation (gyrotrons, backward wave oscillators, helical undulators) the elementary act is based on the radiation from an electron in a circular motion generated by magnetic fields. In this sense, the elementary acts in those sources and in the problem we consider are similar. We show that in the presence of a ball, under certain conditions, the yield from the elementary act can be amplified by an order of magnitude compared with the corresponding radiation in the vacuum. We expect that the amplification effect discussed above will take place in a more general case of helical motion (used in the abovementioned sources of THz radiation). For that motion to increase the effective interaction time of the charge with medium one can consider a dielectric cylinder instead of a ball. It is expected that similar features will be present for the cylindrical case as well. It would be also of interest to consider the radiation for bunches with the length of the order of radiation wavelength and investigate the coherent effects as function of the bunch structure. These points require further investigations and will be discussed elsewhere.

## Acknowledgments

The work was supported by the RA Committee of Science, in the frames of the research project №18T-1C397.